\documentclass[twocolumn,showpacs]{revtex4}
\usepackage{graphicx}
\usepackage{dcolumn}
\usepackage{amsmath}
\makeatletter
\def\btt#1{\texttt{\@backslashchar#1}}
\DeclareRobustCommand\bblash{\btt{\@backslashchar}}
\makeatother

\begin{document}

\title{{Non-marginally bound inhomogeneous dust collapse in higher
dimensional space-time}}

\author{S.G.~Ghosh}
\thanks{E-mail:sgghosh@iucaa.ernet.in}
\affiliation{Department of Mathematics, Science College, Congress Nagar,
Nagpur - 440 012, India}

\author{A.~Banerjee}\thanks{E-mail:asitb@cal3.vsnl.net.in}
\affiliation{Department of  Physics, Jadavpur University,
Calcutta 700 032, India}

\date{\today}

\begin{abstract}
We investigate the occurrence and nature of a naked singularity in
the gravitational collapse of an inhomogeneous dust cloud
described by a self-similar higher dimensional Tolman-Bondi
space-time. Bound, marginally bound and unbound space-times are
analyzed.  The degree of inhomogeneity of the collapsing matter
necessary to form a naked singularity is given.
\end{abstract}


\pacs{04.50.+h, 04.20.Dw, 04.70.Bw, 04.20.Jb}

\maketitle

\section{Introduction}

The Cosmic Censorship conjecture (CCC) of Penrose \cite{rp} states
that in generic situation all singularities arising from regular
initial data are clothed by event horizon and hence invisible to
distant observers. According to strong version of the conjecture,
such singularities are invisible to all observers, i.e., there
occur no naked singularities for any observer. The conjecture
plays a fundamental role in the theory of black holes. Despite the
flurry of activity over years, the validity of this conjecture is
still an open question (see, e.g., \cite{r1} for reviews on
conjecture). However, there exist many exact solutions of Einstein
equations where naked singularities do arise, e.g., naked
singularities occur in Tolman-Bondi space-times where they are
extensively studied \cite{es,rn,lz,jl,dj,jd1,djd}

Comparatively, much less is known about singularity formation in
higher dimensions (HD) that are being currently considered very
important in view of recent developments in string theory and
other field theories.  In string and M-theory, which may provide a
route towards quantum gravity, gravity is truly higher dimensional
interaction, which becomes effective four dimensional (4D) at low
energies.  This provoked great interest among theoretical
physicist in studying physics in HD \cite{hd}.  It would be,
therefore, highly desirable to understand the final fate of
gravitational collapse in HD space-times.  To understand the
general collapse problem as well as nature of singularities, one
would like to analyze exact solutions, preferably in close form,
of HD Einstein equations. However the non-linearity of field
equations makes them difficult, even in spherical symmetry.

A model in which analytical treatment appears feasible is that of
5D Tolman-Model that describes gravitational collapse of
spherically symmetric inhomogeneous dust in a 5D space-time, since
in this case the general exact solutions, for both marginally
bound ($f=0$) and non-marginally bound ($f \neq 0$), are known in
close form \cite{bc}.  This is unlike the 4D case, where the
cooresponding solution is available in parametric form only.
Hence, we shall restrict  ourselves  to the 5D non-marginally
bound collapse. Further, the 5D space-time is particularly more
relevant because both 10D and 11D supergravity theories yield
solutions where a 5D space-time results after dimensional
reduction \cite{js}.

The objective of this paper is to analyze  the collapse of an
inhomogeneous dust cloud in a 5D self-similar space-time and
extend the earlier results \cite{gb,bsc} beyond marginally bound
case.  We also access the curvature strength of central shell
focusing singularities.  We find that gravitational collapse of a
self-similar 5D space-time gives rise to a strong-curvature shell
focusing naked singularity, providing an explicit counter-example
to the CCC.

We have used units which fix the speed of light and the
gravitational constant via $8\pi G = c = 1$.

\section{Higher Dimensional Tolman-Bondi Models}
We start with a brief summary of the 5D spherical symmetric
inhomogeneous dust models.  The metric for the 5D case, in
comoving coordinates, assumes the form:
\begin{equation}
ds^2 = - e^{\nu(t,r)} dt^2 + e^{\lambda(t,r)} dr^2 + R^2(t,r) d
\Omega^2 \label{eq:me}
\end{equation}
where $d\Omega^2 = d \theta_1^2+ sin^2 \theta_1 (d
\theta_2^2+sin^2 \theta_2 d\theta_3^2)$ is the metric of a $3$
sphere, $r$ is the comoving radial coordinate, $t$ is the proper
time of freely falling shells, $R$ is a function of $t$ and $r$
with $R \ge 0$ (equality holds at the origin) \cite{bsc}.  Here

\begin{eqnarray}
&&e^{\nu} = 1   \label{ea}
\\
&&e^{\lambda} = \frac{R{'}^2}{1+f(r)} \label{eb}
\end{eqnarray}
where the  prime denotes a partial derivative with respect to $r$,
$f(r)$ is an arbitrary function satisfying
\begin{equation}
f(r) > -1
\end{equation}
and $R'$ is assumed to be positive, in order to avoid situation
where negative mass shells are present \cite{ms}.  The function
$R(t,r)$ is the solution of
\begin{equation}
\dot{R}^2 = \frac{F(r)}{R^2} + f(r) \label{eq:fe}
\end{equation}
where the over-dot denotes the partial derivative t.  Since in the
present discussion we are concerned with collapse, we require that
$\dot{R}(t,r) < 0$. The functions $F(r)$ is another arbitrary
function and like $f(r)$, it also results from the integration of
the field equations. The function $F$ must be positive, because
$F< 0$ implies the existence of negative mass. This can be seen
from the mass function $m(t,r)$ \cite{ms},
 which in the 5D Tolman-Bondi case is given by
\begin{equation}
m(t,r) = R^2 \left(1 - g^{ab}R,_{a}R,_{b} \right)
\end{equation}
Using Eqs. (\ref{eq:me}) and (\ref{eq:fe}), this implies that
\begin{equation}
m(r) = F(r)
\end{equation}
Equation (\ref{eq:fe}) has three types of solutions, namely,
hyperbolic, parabolic and elliptic solutions, depending on whether
$f(r)>0$, $f(r)=0$ or $f(r)< 0$, respectively.  We note that $F'$
(as well as F) must be positive because mass-energy density
$\epsilon(t,r)$, which is given by
\begin{equation}
\epsilon(t,r) = \frac{3 F'}{2 R^3 R'}  \label{eq:edt}
\end{equation}
must be non-negative, and since we have assumed $R'>0$, it follows
that
 $F'>0$.

Integration Eq. (\ref{eq:fe}) shows that the evolution of the
shell is given by
\begin{equation}
R^2 = \left[f [t_c(r)-t]^2 + 2 \sqrt{F} [t_c(r)-t] \right]
\label{eq:fg}
\end{equation}
and where $t_{c}(r)$ is a function of integration which represents
the time taken by the shell with coordinate $r$  to collapse to
the center. This is a new exact solution of 5D spherical collapse
found, recently, by Banerjee and Chatterjee \cite{bc}. The
solution  (\ref{eq:fg}) is not unique.  In an earlier paper
\cite{bsc} a different solution was given. But, main problem with
that solution is that one cannot smoothly pass over to marginally
bound case when $f = 0$.  The solution  (\ref{eq:fg}) has no such
disadvantage. Hence we have chosen the  solution  (\ref{eq:fg}) to
study nature of singularities for all cases: $f > 0$, $f=0$ as
well as $f < 0$. The  Eq. (\ref{eq:fe}) is also interesting in the
sense that in the analogous 4D case the solutions can't be obtain
in close form \cite{rn,dj}. The three arbitrary functions $F(r)$,
$f(r)$ and $t_c(r)$ completely specify the behavior of shell of
radius $r$.

It is possible to make an arbitrary relabelling of spherical dust
shells by $r \rightarrow g(r)$, without loss of generality, we fix
the labelling by requiring that, on the hypersurface $t = 0$, $r$
coincides with the radius
\begin{equation}
R(0,r) = r              \label{eq:ic}
\end{equation}
This corresponds to the following choice of $t_{c}(r)$
\begin{equation}
t_{c}(r) = \left\{ \begin{array}{ll}
   - \frac{\sqrt{F} \mp \sqrt{F + f r^2}}{f}, &   \mbox{$f \neq 0$}, \\
   + \frac{r^2}{2\sqrt{F}}                &       \mbox{$f = 0 $}.   \\
     \end{array}
        \right.  \label{eq:tc}
\end{equation}
The time coordinate and radial coordinate are respectively in the
ranges $ - \infty < t < t_{c}(r)$ and $0 \leq r < \infty$.

Let $\rho(r)$ be the function given by
\begin{equation}
\rho(r) \equiv \epsilon(0,r) = \frac{3 F'}{2 r^3} \Rightarrow F(r)
= \frac{2}{3} \int \rho(r) r^3 dr  \label{eq:Fr}
\end{equation}
Given a regular initial surface, the time for the occurrence of
the central shell-focusing singularity for the collapse developing
from that surface is reduced as compared to the 5D case for the
marginally bound collapse. The reason for this stems from the form
of the mass function in Eq. (\ref{eq:Fr}). In a ball of radius $0$
to $r$, for any given initial density profile $\rho(r)$, the total
mass contained in the ball is greater
 than in the corresponding 5D case. In the 5D case, the mass function
$F(r)$ involves the integral $\int \rho(r) r^2 dr$ \cite{r1}, as
compared to
 the factor $r^3$ in the 5D case. Hence, there is relatively more
 mass-energy collapsing in the space-time as compared to the 5D case,
 because of the assumed overall positivity of mass-energy (energy condition).
 This explains why the collapse is faster in the 5D case.

In order to study the collapse of a finite spherical body, we have
to match the solution along the time-like surface at some $r =
r_c$ to the 5D Schwarzschild exterior. The 5D metric (\ref{eq:me})
can be matched to 5D Schwarzschild metric
\begin{equation}
ds^2 = - \left(1 - \frac{m_s}{r_s^2} \right) dt^2 +
 \left(1 - \frac{m_s}{r_s^2}\right)^{-1} dr_s^2+ r^2 d \Omega^2 \end{equation}
on a spherical hypersurface $\Sigma$ and the junction conditions
yield $F = m_s$ \cite{bc}.  Here $r_s = R(t,r_c)$ and $m_s$ is the
total mass enclosed within the coordinate radius $r_c$.

Further, 5D Friedmann solutions corresponds to $t_c(r) = 0$, and
with a suitable coordinate choice for r one can set
 $f(r) = -k r^2$, $R(t,r) = r ~S(t)$ and $F(r) = Ar^4$ \cite{bc}.
 Here $k$ and $A$ are constants, and $S(t)$ is the scale factor.
\section{Existence and structure of naked singularities}
It has been shown \cite{rn} that Shell-crossing singularities
(characterized by $R'=0$ and $R>0$) are gravitationally weak and
hence such singularities cannot be considered
 seriously.  Hence, we shall confine our discussion to the
central shell focusing singularity. The Kretschmann scalar ($K =
R_{abcd} R^{abcd}$, $R_{abcd}$ the Riemann tensor).  For the
metric (\ref{eq:me}), it reduces to
\begin{equation}
K = 7 \frac{F{'}^2}{R^6 R{'}^2} - 36 \frac{F F'}{R^7 R'} + 72
\frac{F^2}{R^8}
      \label{eq:ks}
\end{equation}
The Kretschmann scalar and energy density both diverge at
$t=t_{c}(r)$ indicating the presence of a scalar polynomial
curvature singularity \cite{he}. It is known that, depending upon
the inhomogeneity factor, the 5D Tolman-Bondi solutions  admits a
central shell focusing naked singularity in the sense that
outgoing geodesics emanate from the singularity.  Here we wish to
investigate the similar situation in our 5D  space-time.

The self-similar solutions have attracted considerable attraction
in recent decades and  they play a crucial role in many
cosmological and astrophysical contexts \cite{bj}.   Spherically
symmetric self-similar solutions of Einstein equations
 are characterized by the fact that the space-time possess
 a homothetic killing vector.  This means the solution is
 unchanged by transformation $t
\rightarrow at$, $r \rightarrow ar$ for any constant $a$. Thus the
self-similarity demands that

\begin{eqnarray}
&& F(r) = \zeta^2 r^2  \label{ssa} \\
&& f(r) = \mbox{Const.} \label{ssb} \\
&& t_c(r) = B r \label{ssc}
\end{eqnarray}

The parameter $B$ gives the inhomogeneity of the collapse.
  For $B=0$ all shells collapse at the
same time.  For higher $B$ the outer shells collapse much later
than the central shell. From Eqs. (\ref{eq:tc}) and (\ref{ssc}),
we obtain an interesting relation
\begin{eqnarray}
\zeta = \frac{1 - B^2 f}{2 B}
\end{eqnarray}
This expression reduces to $\zeta = 1/2 B$ \cite{bsc,gb} when
$f=0$. We notice that $\zeta > 0$, and so $B^2 f < 1$. We are
interested in the causal structure of the space-time when the
central shell collapses to the center ($R=0$).

From Eqs. (\ref{eq:Fr}) and (\ref{ssa}), the energy density at the
singularity
\begin{equation}
\rho = \frac{3 \zeta^2}{r^2} \label{eds}
\end{equation}
and the equation of general density becomes
\begin{eqnarray}
&& \epsilon ={3 \zeta^2 y}/\Big[t^2 \left[f (B y-1)^2 + 2 \zeta y
(By -1)\right] \nonumber \\
&&\times  \left[f B (B y -1)+\zeta (2 B y -1) \right] \Big]   =
\frac{C(y)}{t^2} \label{ed}
\end{eqnarray}
where $y=t/r$ is usual similarity variable.  Note that $\epsilon$
is singular at $t = B r$. The nature (a naked singularity or a
black hole) of the singularity can be characterized by the
existence of radial null geodesics emerging from the singularity.
The singularity is at least locally naked if there exist such
geodesics, and if no such geodesics exist, it is a black hole. The
critical direction is the Cauchy horizon. This is the first
outgoing null geodesic emanating from $r=t=0$. The Cauchy horizon
of the self-similar space-time has $y = r/t$ = const \cite{lz,jl}.
The equation for outgoing null geodesics is
\begin{eqnarray}
\frac{dt}{dr} = \frac{R'}{\sqrt{1+f}}
\end{eqnarray}
Hence along the Cauchy horizon, we have
\begin{eqnarray}
y^2 R{'}^2 = 1+f \label{ch}
\end{eqnarray}
and using Eqs. (\ref{ch}) and (\ref{eq:fg}), with our choice of
the scale,
 we obtain the following algebraic equation:
\begin{eqnarray}
(1+f) \left[f (B y - 1)^2 + 2 \zeta y (B y -1) \right] \nonumber
\\ - y^2 \left[f B (B y - 1) + \zeta (2 B y - 1) \right]^2 = 0
\label{eq:ae}
\end{eqnarray}
This algebraic equation governs the behavior of the tangent vector
near the  singular point. The central shell focusing
 singularity is at least locally naked, if Eq. (\ref{eq:ae})
 admits one or more positive real roots. Hence in the absence of
positive real roots, the collapse will always lead to a black
hole. Thus, the occurrence of positive real roots implies that the
strong CCC is violated, though  not necessarily the weak CCC. If
Eq. (\ref{eq:ae}) has only one positive root, a single radial null
 geodesic would escape from the singularity, which amounts to
 a single wave front being emitted from the singularity
 and hence singularity would appear to be naked only, for an instant, to
an asymptotic observer.  A naked singularity forming in
gravitational collapse could be physically significant if it is
visible for a finite period of time, to an asymptotic  observer,
i.e., a family of geodesics must escape from the singularity. This
happens only when  Eq. (\ref{eq:ae}) admits at least two positive
real roots \cite{r1}.  In this paper we are concerned only with
such singularities.  Hence we shall look for conditions
 for which  Eq. (\ref{eq:ae}) admits at least two positive real roots.

We start from the marginally bound case, which has been analyzed
earlier \cite{bsc,gb}. We already know what happens and we refer
the reader to these papers for details. Setting $f=0$, Eq.
(\ref{eq:ae}) simplifies to
\begin{equation}
y \zeta^2 (2 B y -1)^2 = 2 \zeta (B y -1) \label{ae2}
\end{equation}
To facilitate comparison with the work of Ghosh and Beesham
\cite{gb}, we introduce a new variable $y = 1/X$ and use the value
of $\zeta$ and  after some rearrangement, Eq. (\ref{ae2}) takes
the same form as in  \cite{gb}:
\begin{equation}
X^2 \left[1 - \frac{X}{B} \right] = \left[1 - \frac{1}{2 B} X
\right]^2 \label{eq:pe5}
\end{equation}
It can be shown that Eq. (\ref{eq:pe5}) has two positive roots if
$B > B_c = 1.6651$. This is slightly higher than the analogous
value, $B_c^4 = 1.56736$, in 5D.  The corresponding Cauchy horizon
evolves as $y = 0.78611$.   Thus in 5D one needs higher
inhomogeneity to produce naked singularity.  For $B > B_c$, two
solutions exist, the largest $y$ gives the Cauchy horizon. Other
solution is termed as self-similar horizon \cite{jl} (see Table
I).

It is seen that Eq. (\ref{eq:ae}) admit two positive roots for $y$
when $\zeta = \zeta_c  \leq 0.30028 $ and hence  referring to our
above discussion singularities are naked for $\zeta \in (0,
\zeta_c$ and black holes form otherwise.  Thus $\zeta _ c$ is the
value of $\zeta$ where a transition from naked singularities to
black holes occur. The quantity $\zeta_c$ is called critical
parameter.  It is interesting to see that the value of $\zeta_c$
decreases in 5D and hence one can say that naked singularity
spectrum, of the 4D dust collapse, gets partially covered in 5D.
Our results are in agreement with earlier results.  A similar
situation also occur in higher dimensional radiation collapse
\cite{ns}.

Next we compare the behavior found in the marginally bound ($f=0$)
case with that found in non-marginally bound case. In order to
investigate the changes introduced, at least qualitatively in the
above picture when $f \neq 0$, we have solved (\ref{eq:ae})
numerically and summarize the results in table II. For negative
values of $f$, there is critical value, $B_c$, such that for $B
\geq B_c$ collapse always leads to a visible singularity and
covered otherwise. The numerical results show that for larger
(smaller) positive curvature, i.e., for larger(smaller) magnitude
of $f$  one needs  a collapse with higher (lower) inhomogeneity to
produce  a naked singularity.

Stars are bound objects but an unbound state can occur, e.g., if
there is collision of two stars.  If $f > 0$, we observe that
collapse lead to a visible singularity for $B_c \leq B \leq B_u$,
where $B_u$ is some upper bound for $B$.  In this case, there are
three solutions (see Table II).  Further, it is interesting to
note that $\zeta$ and $B_C$ decreases
 as we increase the value of $f$.  Thus for larger
(smaller) negative curvature  one needs  a collapse with lower
(higher) inhomogeneity to produce a naked singularity.

Finally, the apparent horizon is formed when the boundary of
trapped three spheres are formed.  The apparent horizon is the
solution of
\begin{equation}
g^{ab}R_{,a}R_{,b} = \dot{R}^2 - \frac{R{'}^2}{e^{\lambda}}
\label{ah2}
\end{equation}
Considering Eqs. (\ref{eb}) and (\ref{eq:fe}), we have the
apparent horizon in the interior dust ball lies at $R^2 = F$. The
corresponding time $t_{\mbox{ah}}(r)$ is given by
\begin{equation}
t_{\mbox{ah}}(r) = t_{c}(r) - \frac{1}{2} \sqrt{F} \label{ah}
\end{equation}
and because of Eqs. (\ref{ssc}), (\ref{eq:tc}) and (\ref{ah}):
$t_{\mbox{ah}}(r)- t_{\mbox{Cauchy}}(r) < 0$. Thus apparent
horizon always appear before the Cauchy horizon has formed and
 singularity is globally naked. The global nakedness of the
singularity can then be seen by making a junction onto 5D
Schwarzschild space-time \cite{lz,jl}.

\begin{table}
\caption{Variation of $y$, $\lambda$ for different $B$ ($f = 0$)}
\label{table2}
\begin{ruledtabular}
\begin{tabular}{ccc}
$B$ & $\lambda$ & $y$  \\
\colrule
  1.7 & 0.294118    & 0.714512, \; 0.846397 \\
  2  & 0.25         & 0.549139, \; 0.929956 \\
  2.5 & 0.2         & 0.420428, \; 0.965243 \\
  3  & 0.166667     & 0.344179, \; 0.978711 \\
  3.5 & 0.142857    & 0.292235, \; 0.985525 \\
\end{tabular}
\end{ruledtabular}
\end{table}

\begin{table}
\caption{Variation of $B_C$ $\lambda$ and $y_0$ for various
$f$} \label{table1}
\begin{ruledtabular}
\begin{tabular}{cccc}
$f$ &  $B_C$ & $\lambda$ & Equal roots  $y_0$ \\
\colrule
 - 0.4  & 4.16347     & 0.952787 &  0.3978            \\
 - 0.3  & 2.82096     & 0.600388 &  0.53331           \\
 - 0.2  & 2.23607     & 0.447214 &  0.63245 \\
    0   & 1.6651      & 0.300283 &  0.78611 \\
   0.4  & 1.18805     & 0.183246 &  1.01733 \\
   1    & 0.890515    & 0.116216 &  1.28286 \\
   2    & 0.671855    & 0.0723525 & 1.62888 \\
   3    & 0.560097    & 0.0525573 & 1.91208 \\
\end{tabular}
\end{ruledtabular}
\end{table}

\subsection{Strength of Naked Singularity}
An important aspect of a singularity is its gravitational strength
\cite{ft1}.  There have been attempts to relate it to stability
\cite{djd}.  A singularity is gravitationally strong or simply
strong if it destroys by crushing or stretching any object which
fall into it.  It is widely believed
 that a space-time does not admit an extension through a singularity if it is
a strong curvature singularity in the sense of Tipler \cite{ft}.
Clarke and Kr\'{o}lak \cite{ck} have shown that a sufficient
condition for a strong curvature singularity as defined by Tipler
\cite{ft} is that for at least one non-space like geodesic with
affine parameter $k$, in the limiting approach to the singularity,
we must have
\begin{equation}
\lim_{k\rightarrow 0}k^2 \psi = \lim_{k\rightarrow 0}k^2 R_{ab}
K^{a}K^{b} > 0 \label{eq:sc}
\end{equation}
where $R_{ab}$ is the Ricci tensor. Our purpose here is to
investigate the above condition along future directed radial null
geodesics, which emanate from the naked singularity. Now $k^2
\psi$, with the help of Einstein equations and Eq. (\ref{ed}) can
be expressed as
\begin{equation}
k^2 \,\psi = C \left[\frac{k}{t} \frac{dt}{dk}\right]^2
\label{eq:sc1}
\end{equation}
Clearly the null geodesic equations for Cauchy horizon give $t =
k^{\alpha}$ \cite{lz}. We find that
\begin{equation}
\lim_{k\rightarrow 0}k^2\, \psi = C_0 \,\alpha^2 >0
\end{equation}
where $\lim C = C_0$.  Thus, along Cauchy horizon, the strong
curvature condition is satisfied.

\section{Concluding remarks}
The Tolman-Bondi metric has been widely used, in 4D case, to
understand the final fate of gravitational collapse. It has been
found that both naked singularities and black holes form depending
upon the choice of initial data.  Indeed, one can safely assert
that end state of 4D Tolman-Bondi collapse is now completely known
in dependence of choice of initial data.   In the absence of the
proof of either version of CCC, it was used as tool to get
insights in to  more general gravitational collapse situations. In
this paper we have investigated the influence of extra dimensions
on Tolman-Bondi dust collapse. The occurrence and curvature
strength of a shell focusing naked singularity in a 5D
self-similar spherically symmetric collapse of a dust cloud has
been investigated.
  We found that the
scenario changes in many ways. The extra dimensions is a shrinkage
of the naked singularity initial data space, or an enlargement of
the black hole initial data space of 4D Tolman-Bondi collapse.  We
found that naked singularities in our case develop for a slightly
higher value of the inhomogeneity parameter in comparison to the
analogous situation in the 4D case.  Thus the presence of an extra
dimension does alter the established picture of 4D Tolman-Bondi
collapse, but cannot completely cover the naked singularities.
Further, Along the null ray emanating from the naked singularity,
the strong curvature condition (\ref{eq:sc}) is satisfied.  The
formation of these naked singularities violates the CCC.

The crucial issue associated with naked singularities is the
question of their stability.  The main point is that if  naked
singularities are not stable or generic in some suitable sense,
then they may not be physically realizable. However, the problem
is that we do not have well defined criteria of stability. Under
such circumstances, one can consider a perturbation of currently
available collapse scenarios and examine if the singularity still
persists or gets covered. It is worth noting that, by introduction
of extra dimensions, there exist a set of nonzero measure of
values of $\zeta$ for which strong curvature naked singularities
develop, and the CCC is violated. Thus presence of extra
dimensions do not remove naked singularity of  4D Tolman-Bondi
collapse. As a result, one can say that the naked singularity of
4D Tolman-Bondi collapse is stable to the introduction extra
dimension.

The exact analysis discussed, for non-marginally bound case, is
specialized for the 5D Tolman-Bondi models and can't be done even
in analogous 4D case.

In conclusion, we have studied the development of a strong
curvature naked singularity in 5D Tolman-Bondi collapse that
violates the strong CCC. The Kretschmann scalar diverges in the
approach to the singularity. Thus the naked singularity in 5D
Tolman-Bondi collapse  could be considered as a physically
significant curvature singularity.

\acknowledgments 
Authors  would like to thank  IUCAA, Pune for a visit under a
ssociate-ship programme where this work was done.

\noindent

\small

\begin{thebibliography}{99}
\bibitem{rp} R. Penrose, {\it   Riv. Nuovo Cimento} {\bf 1}, 252
 (1969); in {General Relativity, an Einstein Centenary
 Volume}, ed.  S. W.Hawking and W. Israel (Cambridge
 University Press, Cambridge, England, 1979).
\bibitem{r1} P. S. Joshi, {\it   Global Aspects in Gravitation and
Cosmology} (Clendron, Oxford, 1993);
 C.J.S. Clarke, {\it  Class. Quantum Grav.} {\bf 10}, 1375 (1993);  R.M. Wald,
gr-qc/9710068;  S. Jhingan and G. Magli, gr-qc/9903103; T.P.
Singh, {\it  J. Astrophys. Astron.} {\bf 20}, 221 (1999); P.S.
Joshi, {\it  Pramana} {\bf 55}, 529 (2000).
\bibitem{es} D.M. Eardley and L. Smarr, {\it  Phys. Rev.}
{\bf D 19}, 2239 (1979); D.M. Eardley, {\it  Gravitation in
Astrophysics, ASI Series}, edited by B. Carter and J.B. Hartle
(NATO Advanced Study Institute, Series B: Physics, Vol 156)(Plenum
Press, New York, 1986) pp 229-235.
\bibitem{rn} R.P.A.C. Newman, {\it  Class. Quantum Grav.} {\bf 3}, 527 (1986).
\bibitem{lz}B. Waugh and K. Lake, {\it  Phys. Rev.} {\bf D 38}, 1315 (1988).
\bibitem{jl} J.P.S. Lemos, {\it  Phys. Lett. }
{\bf A 158}, 271 (1991); {\it  Phys. Rev. Lett.} {\bf 68}, 1447
(1992).
\bibitem{dj} I.H. Dwivedi and P.S. Joshi, {\it  Class. Quantum Grav.}
 {\bf 9}, L69 (1992).
 P.S. Joshi and T.P. Singh, {\it  Gen. Relativ. Gravit.}
 {\bf 27}, 921 (1995); {\it  Phys. Rev. } {\bf D 51}, 6778 (1995).
\bibitem{jd1}P.S. Joshi and I.H. Dwivedi, {\it  Phys. Rev.}
 {\bf D 47}, 5357 (1993).
\bibitem{djd} S.S. Deshingkar, P.S. Joshi, and I.H. Dwivedi,
{\it  Phys. Rev.} {\bf D 59}, 044018 (1999).
\bibitem{hd} J. Soda and K. Hirata, {\it  Phys. Lett.} {\bf B 387}, 271 (1996);
  A. Ilha and J.P.S. Lemos,  {\it  Phys. Rev.} {\bf D 55}, 1788
(1997);   A. Ilha, A. Kleber and J.P.S. Lemos, {\it  J. Math.
Phys.} {\bf 40}, 3509 (1999);  A. V. Frolov {\it  Class. Quantum
Grav.} {\bf 16}, 407 (1999);  J.F.V. Rocha and  A. Wang {\it
ibid.} {\bf 17}, 2589 (2000).
\bibitem{bc}  A. Banerjee and S. Chatterjee, {\it   Pre-print} (2002).
\bibitem{js} J.J. Schwarz, {\it  Nucl. Phys.} {\bf B226}, 269 (1983).
\bibitem{gb} S.G. Ghosh and A. Beesham {\it  Phys. Rev.} {\bf D 64},
124005 (2001).
\bibitem{bsc} A. Banerjee, A. Sil and S.  Chatterjee {\it  Astrophys. J.}
 {\bf 422}, 681 (1994); A. Sil and S. Chatterjee {\it  Gen. Relativ. Gravit.}
 {\bf 26}, 999 (1994).
\bibitem{bj}  B.J. Carr and A.A. Coley,  {\it  Class. Quantum Grav.}
 {\bf 16}, R31 (1999).
 \bibitem{ms} C.W. Misner and D. Sharp {\it  Phys. Rev.} {\bf 136},
b571 (1964).
\bibitem{he} S.W. Hawking and G.F.R. Ellis, {\it   The Large Scale Structure
of Space-time} (Cambridge University Press, Cambridge, 1973).
\bibitem{ns} S.G. Ghosh and N. Dadhich {\it  Phys. Rev. } {\bf D 64},
047501 (2001).
\bibitem{ft1} F.J. Tipler {\it  Phys. Lett. } {\bf A 64},8 (1987).
\bibitem{ft} F.J. Tipler, C.J.S. Clarke, and G.F.R. Ellis in
{\it   General Relativity and Gravitation}, edited by A Held
(Plenum, New York, 1980).
\bibitem{ck} C.J.S. Clarke and  A.  Kr\'{o}lak {\it  J. Geom. Phys.} {\bf
2}, 127 (1986).
\end{thebibliography}
\end{document}